 \documentstyle[11pt,aaspp4,psfig]{article}

 \slugcomment{submitted to \it The Astrophysical Journal Supplement Series}

 \lefthead{Montes et al.}
 \righthead{Library of medium-resolution FOE echelle spectra}
 
\begin{document}
 
 \title {Library of medium-resolution fiber optic echelle spectra
  of F, G, K, and M field dwarfs to giants stars
 }
 
 \author{David Montes\altaffilmark{1,2}, Lawrence W. Ramsey\altaffilmark{1,4},
 Alan D. Welty\altaffilmark{3,4}}
 
 \altaffiltext{1}{The Pennsylvania State University,
 Department of Astronomy and Astrophysics,
 525 Davey Laboratory, University Park, PA 16802, USA
 (dmg@astro.psu.edu, lwr@astro.psu.edu)}
 
 \altaffiltext{2}{Departamento de Astrof\'{\i}sica,
 Facultad de F\'{\i}sicas,
  Universidad Complutense de Madrid, E-28040 Madrid, Spain
 (dmg@astrax.fis.ucm.es)}
 
 \altaffiltext{3}{Space Telescope Science Institute,
 3700 San Martin Drive, Baltimore, MD 21218, USA 
 (welty@stsci.edu)}
 
 \altaffiltext{4}{Visiting Astronomer, Kitt Peak National Observatory,
 National Optical Astronomy Observatories, which is operated by the
 Association of Universities for Research in Astronomy, Inc. (AURA),
 under cooperative agreement with the National Science Foundation}

 \begin{abstract}
 
 We present a library of Penn State Fiber Optic Echelle (FOE) observations
 of a sample of field stars with spectral types F to M and luminosity classes
 V to I.
 The spectral coverage is from 3800~\AA$\ $ to 10000~\AA\ with 
 nominal a resolving power 12000.
 These spectra include many of
 the spectral lines most widely used as
 optical and near-infrared indicators of chromospheric activity
 such as the Balmer lines (H$\alpha$ to H$\epsilon$),
 Ca~{\sc ii} H \& K, 
 Mg~{\sc i} b triplet,
 Na~{\sc i} D$_{1}$, D$_{2}$, He~{\sc i} D$_{3}$,
  and Ca~{\sc ii} IRT lines.  There are also a large number of
 photospheric lines, which can also be affected by chromospheric activity,
 and temperature sensitive photospheric features such
 as TiO bands. 
 The spectra have been compiled with the goal of providing a set of
 standards observed at medium resolution.
 We have extensively used such data for the study of active chromosphere
stars by
 applying a spectral subtraction technique.
 However, the data set presented here can also be utilized in a wide
variety of 
 ways ranging from radial velocity templates to study of variable stars
 and stellar population synthesis.
 This library can also be used for spectral classification purposes
 and determination of atmospheric parameters (T$_{\rm eff}$, $\log{g}$,
 [Fe/H]).
 A digital version of all the fully reduced spectra 
 is available via ftp and the
 World Wide Web (WWW) in FITS format.

 
 \end{abstract}
 
 \clearpage
 
 \section{Introduction}
 
 Spectral libraries of late-type stars
 with medium to high resolution and large spectral coverage
 are an essential tool for the study of the chromospheric activity in
 multiwavelength optical  observations using the spectral subtraction technique
 (see Barden 1985; Huenemoerder \& Ramsey 1987; Hall \& Ramsey 1992; 
 Montes et al. 1995a, b, c, 1996a, b, 1997b, 1998).
 Furthermore, these libraries are also very useful in many
 areas of astrophysics such as the stellar spectral classification,
 determination of atmospheric parameters (T$_{\rm eff}$, $\log{g}$,
 [Fe/H]), modeling stellar atmospheres,
 spectral synthesis applied to composite systems,
 and spectral synthesis of the stellar population of galaxies.
 
 In previous work Montes et al. (1997a, hereafter Paper I) presented 
 a library of high and mid-resolution (3 to 0.2 \AA) spectra in the
 Ca~{\sc ii} H \& K, H$\alpha$, H$\beta$,
 Na~{\sc i} D$_{1}$, D$_{2}$, and He~{\sc i} D$_{3}$
 line regions of F, G, K, and M field stars.
 A library of echelle spectra
 of a sample of F, G, K, and M field dwarf stars is presented in
 Montes \& Mart\'{\i}n (1998, hereafter Paper II)
 which is an extension of Paper I 
 to higher spectral resolution (0.19 to 0.09 \AA)
 covering a large spectral range (4800 to 10600 \AA).
 
 The spectral library presented here expands upon the data set
 in Papers I and II.
 This library consists of echelle spectra
 of a sample of F, G, K, and M field stars, mainly dwarfs (V), subgiant (IV),
 and giants (III) but also some supergiants (II, I). 
 The spectral resolving power is intermediate, 
 nominally R~=~12000 ($\approx$~0.5~\AA\ in H$\alpha$),
 but the spectra have a nearly complete optical region coverage
 (from 3900 to 9000\AA).
 These regions includes most of
 the spectral lines widely used as
 optical and near-infrared indicators of chromospheric activity
 such as the Balmer lines (H$\alpha$ to H$\epsilon$),
 Ca~{\sc ii} H \& K,
 Mg~{\sc i} b triplet,
 Na~{\sc i} D$_{1}$, D$_{2}$, He~{\sc i} D$_{3}$,
  and Ca~{\sc ii} IRT lines, as well as 
 temperature sensitive photospheric features such
 as TiO bands.

Recently, Pickles (1998) has taken available published spectra and combined 
them into a uniform stellar spectral flux library.
This library have a wide wavelength, spectral type, and luminosity class
coverage, but a low spectral resolution (R~=~500) and their main purpose is
the synthesis and modeling of the integrated light from composite populations.
However, for other purposes as detailed studies of chromospheric activity, 
 stellar spectral classification, and
 determination of atmospheric parameters,
libraries of higher resolution, as the presented in Paper I and II, 
 Soubiran, Katz, \& Cayrel (1998), and the library presented here  
are needed.
 
 In Sect.~2 we report the details of our observations and data reduction.
 The library is presented in Sect.~3.

 \section{Observations and data reduction}
 
 The echelle spectra presented here
 were obtained during several observing runs 
 with the Penn State Fiber Optic Echelle (FOE)
 at the 0.9-m and 2.1-m telescopes of
 the Kitt Peak National Observatory (KPNO).
 The FOE is a fiber fed prism cross-dispersed echelle medium resolution
spectrograph and is  described in more detail in Ramsey \& Huenemoerder
(1986). It was designed specifically to  obtain in a single exposure a
wide spectral range encompassing all the visible chromospheric  activity
sensitive features. Typical data and performance of the FOE for the
different observing  runs are discussed in 
 Ramsey et al.~(1987); Huenemoerder, Buzasi, \& Ramsey (1989);
 Newmark et al. (1990); Hall et al. (1990);
 Buzasi, Huenemoerder, \& Ramsey (1991); Hall \& Ramsey (1992); Welty \&
 Ramsey (1995); and  Welty (1995).
 
 In Table \ref{tab:obs} we give a summary of observations.
 For each observing run we list the date, the CCD detector used,
  the number of echelle orders included,
 the wavelength range covered ($\lambda$$_{i}$-$\lambda$$_{f}$) and the range
 of reciprocal dispersion achieved (\AA/pixel) from the first to the last
 echelle orders.
 The \AA/pixel value for each order can be found in the header of the spectra.
 The spectral resolution, determined by the FWHM of the arc
 comparison lines, ranges from 2.0 to 2.2 pixels.
 The signal to noise ratio is larger than 100 in all cases.
 Tables~\ref{tab:orders} 
  gives for each observing run the
 spectral lines of interest in each echelle order.
 
 The spectra have been extracted using the standard
 reduction procedures in the 
 IRAF\footnote{IRAF is distributed by the National Optical Observatory,
 which is operated by the Association of Universities for Research in
 Astronomy, Inc., under contract with the National Science Foundation.}
  package (bias subtraction,
 flat-field division, and optimal extraction of the spectra).
 The wavelength calibration was obtained from concurrent spectra 
  of a Th-Ar hollow cathode lamp.
 Finally, the spectra have been normalized by
 a polynomial fit to the observed continuum.

 \section{The library}

 As in Papers I and II, the stars included in the library
 have been selected as stars with low levels of chromospheric activity,
 that is to say,
 stars that do not present any evidence of emission in the core
 of Ca~{\sc ii} H \& K lines in our spectra
 (Montes et al. 1995c, 1996a),
 stars with the lower Ca~{\sc ii} H \& K spectrophometric index S
 (Duncan et al. 1991; Baliunas et al. 1995),
 or stars known to be inactive and slowly rotating stars
 from other sources
 (see Strassmeier et al. 1990; Strassmeier \& Fekel 1990; Hall \& Ramsey
1992).
 
 Table~\ref{tab:par} presents information about the observed stars.
 In this table we give the HD, HR and GJ numbers, name,
 spectral type and luminosity class (T$_{\rm sp}$),
 from the Bright Star Catalogue
 (Hoffleit \& Jaschek 1982; Hoffleit \& Warren 1991),
 the Catalogue of Nearby Stars (Gliese \& Jahreiss 1991),
 and Keenan \& McNeil (1989).
 The exception is some of the M dwarfs for which we list the more recent
 spectral type determination given by
 Henry, Kirkpatrick, \& Simons (1994).
 In column (6) MK indicates if the star is a 
 Morgan and Keenan (MK) Standard Star from Garc\'\i a (1989)
 and Keenan \& McNeil (1989).
 MK* indicates if the star is included in the list of 
 Anchor Points for the MK System compiled by Garrison (1994).
 Column (7) give the metallicity [Fe/H] from Taylor (1994; 1995)
 or Cayrel de Strobel (1992; 1997)
  and column (8) rotational period (P$_{\rm rot}$) and {\it v}~sin{\it i}
 from Donahue (1993), Baliunas et al. (1995),
 Fekel (1997), and Delfosse et al. (1998). 
 We also give, in column (9), the Ca~{\sc ii} H \& K spectrophometric index S
 from Baliunas et al. (1995) and Duncan et al. (1991).
 In column (10) we list information about the observing run in which each
 star have been observed, using a code given
 in the first column of Table~\ref{tab:obs}, 
 the number between brackets give the number of spectra available.
 The last two columns indicate if the star was also included in Papers I
and II.

\begin{figure*}
{\psfig{figure=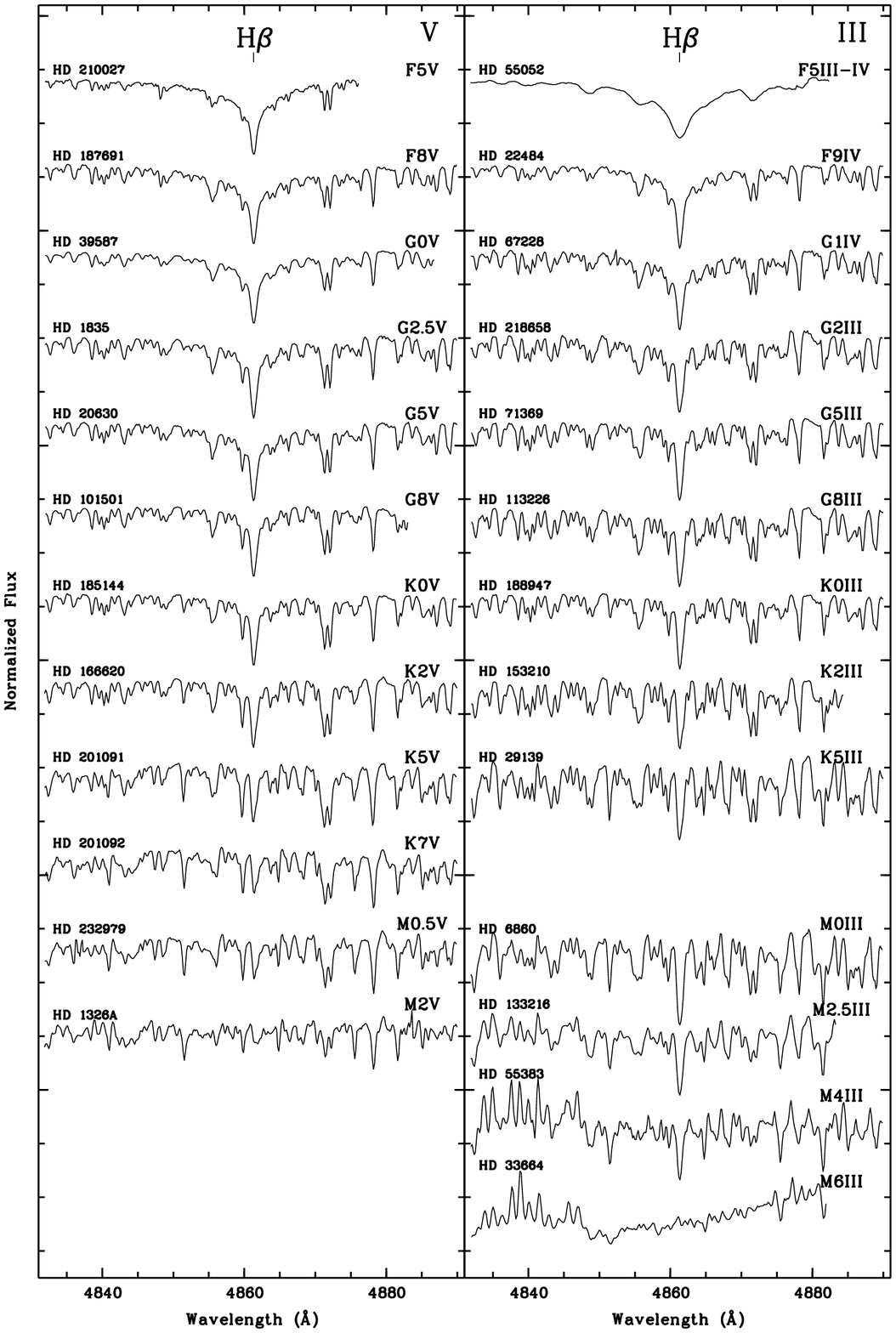,bbllx=47pt,bblly=32pt,bburx=542pt,bbury=767pt,height=22.5cm,width=18.0cm,clip=}}
\caption[ ]{Spectra in the H$\beta$ line region of star with representative
spectral types. Main sequence stars (V) are plotted in the left panel, 
and giants stars (III) in the right panel.
\label{fig:hb} }
\end{figure*}

\begin{figure*}
{\psfig{figure=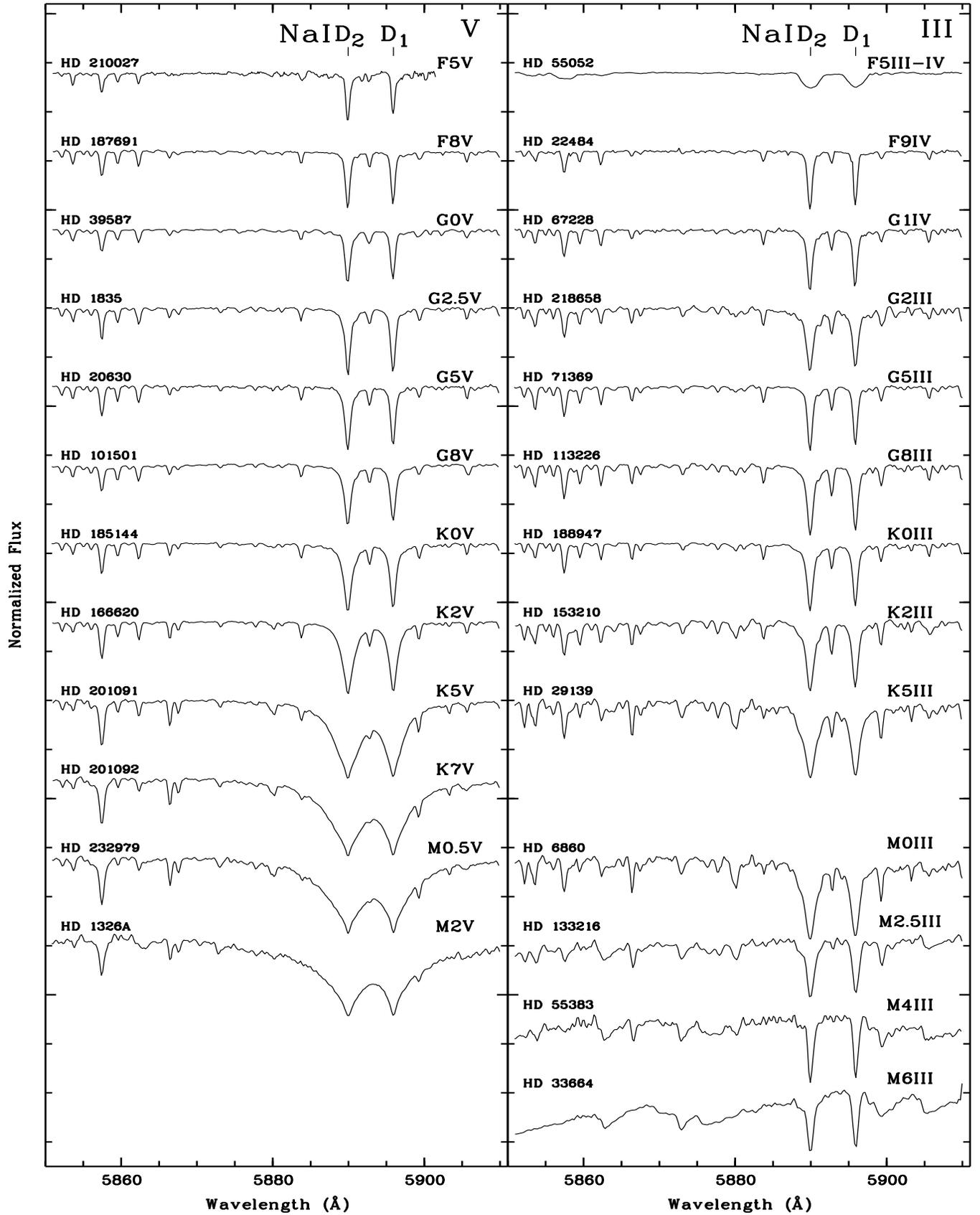,bbllx=47pt,bblly=32pt,bburx=542pt,bbury=767pt,height=22.5cm,width=18.0cm,clip=}}
\caption[ ]{As Fig. 1 in the Na~{\sc i} D$_{1}$, D$_{2}$, He~{\sc i} D$_{3}$ 
line region. 
\label{fig:na} }
\end{figure*}

\begin{figure*}
{\psfig{figure=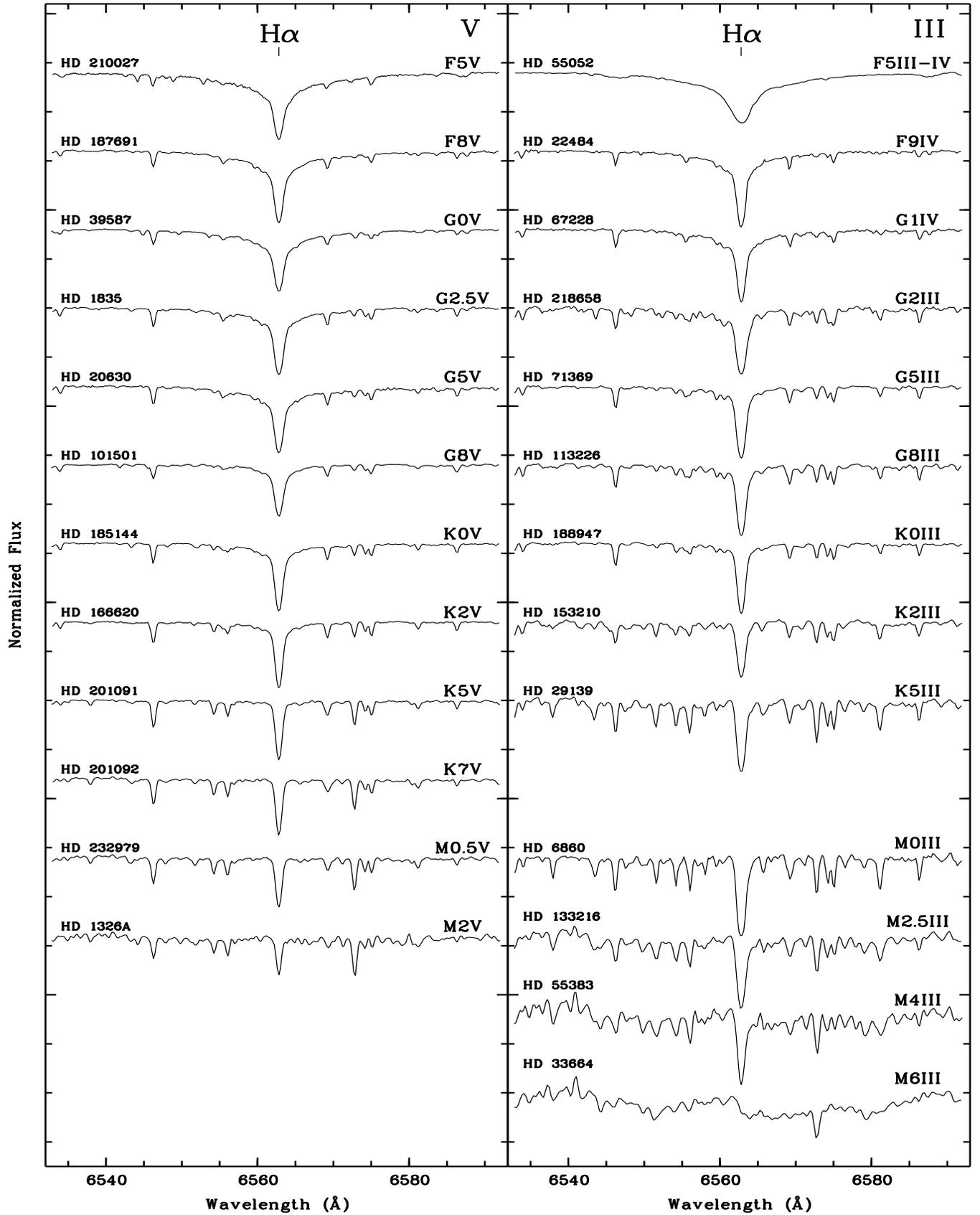,bbllx=47pt,bblly=32pt,bburx=542pt,bbury=767pt,height=22.5cm,width=18.0cm,clip=}}
\caption[ ]{As Fig. 1 in the H$\alpha$ line region.
\label{fig:ha} }
\end{figure*}

\begin{figure*}
{\psfig{figure=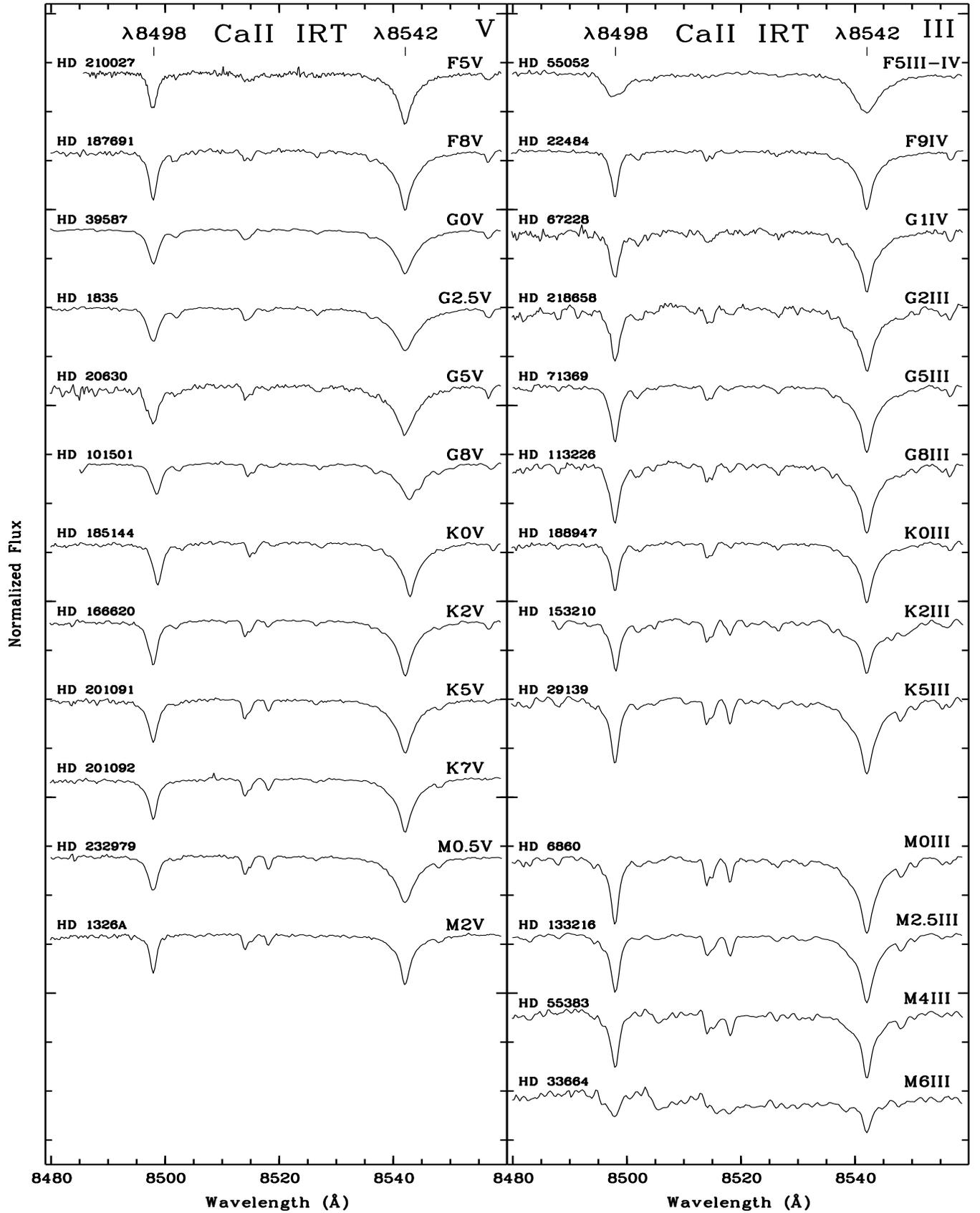,bbllx=47pt,bblly=32pt,bburx=542pt,bbury=767pt,height=22.5cm,width=18.0cm,clip=}}
\caption[ ]{As Fig. 1 in the  Ca~{\sc ii} IRT ($\lambda$ 8498, 8542) 
line region.
\label{fig:cairt} }
\end{figure*}

Representative spectra (from F to M, dwarfs and giants stars)
in different spectral regions are plotted
in figures (\ref{fig:hb} to \ref{fig:cairt})
in order to show the behaviour
of the more remarkable spectroscopic features
with the spectral type and luminosity class.
In order of increasing wavelength we have plotted
the following line regions:
H$\beta$ (Fig.~\ref{fig:hb}),
Na~{\sc i} D$_{1}$, D$_{2}$, and
He~{\sc i} D$_{3}$) (Fig.~\ref{fig:na}),
H$\alpha$ (Fig.~\ref{fig:ha}),and
Ca~{\sc ii} IRT $\lambda$8498, 8542  (Fig.~\ref{fig:cairt}).
In each figure we have plotted main sequence stars 
(luminosity class V) in the left panel,
and giants stars (III) in the right panel.

 A total of 130 stars are included in this library. 
 Many of them have been observed in 
 several observing runs, and in some cases several nights during the same 
 observing run being the total number of spectra 345.
 Using these spectra as well as those of Papers I and II 
 a study of possible short and long term spectroscopic variability
 of some of the multiply observed stars is possible.
 
 A description of the spectral lines most widely used as
 optical and near-infrared indicators of chromospheric activity, as well as
 other interesting spectral lines and molecular bands
 present in the spectral range covered by the spectra 
 can be found in Papers I and II and references therein.
 
 As an illustration of the use of these spectra and those of Papers I and II 
  we intend to analyze temperature sensitive lines
 in order to improve the actual line-depth ratio temperature calibrations
 (Gray \& Johanson 1991, Gray 1994)
 and spectral-class/temperature classifications
 (Strassmeier \& Fekel 1990),
 as well as the determination of fundamental atmospheric 
 parameters T$_{\rm eff}$, $\log{g}$, [Fe/H] 
 (Katz et al. 1998 and Soubiran et al. 1998).  This 
 will be the subject of forthcoming papers.
 
 In order to enable other investigators to make use of the spectra in this
 library for their own purposes, all the final reduced 
 (flattened and wavelength calibrated) multidimensional
 spectra containing all the echelle orders of the stars listed
 in Table~\ref{tab:par} are available
 at the CDS in Strasbourg, France,
 via anonymous ftp to cdsarc.u-strasbg.fr (130.79.128.5).
 They are also available via the World Wide Web at:
 \newline
 {\small http://www.ucm.es/info/Astrof/fgkmsl/FOEfgkmsl.html}.
 \newline
 The data are in FITS format with pertinent header information included 
 for each image.
 In order to further facilitate the use of this library
 one dimensional normalized and wavelength calibrated
 spectra, for the orders
 containing the more remarkable spectroscopic
 features, are also available as separate FITS format files.
 
 In addition this library as well as the libraries presented in Papers I
 and II will be included in the {\it Virtual Observatory}
 (see {\small http://herbie.ucolick.org/vo/}). 
 This is a project to establish
 a new spectroscopic database which will be contained digitized spectra
 of spectroscopic plates as well as spectra observed digitally from
 different observatories. {\it Virtual Observatory} is an 
 International Astronomical Union (IAU) initiative
 through its Working Group for Spectroscopic Data Archives.

 \begin{acknowledgements}
 This research has made use of the SIMBAD data base, operated at CDS,
 Strasbourg, France.
 This work has been supported by the Universidad Complutense de Madrid
 and the Spanish Direcci\'{o}n General de Ense\~{n}anza Superior e 
 Investigaci\'{o}na Cient\'{\i}fica (DGESIC) under grant PB97-0259
 and by National Science Foundation (NSF) grant AST~92-18008.  We also
 acknowledge, with gratitude, KPNO supporting the FOE presence from 1987
 until 1996.
 \end{acknowledgements}




 \begin{table*}
 \caption[ ]{Summary of Observations \label{tab:obs} }
 \small
 \begin{flushleft}
 \begin{tabular}{c l c c c c c c c c c c}
 \noalign{\smallskip}
 \tableline
 \tableline
 \noalign{\smallskip}
  O & Date & CCD Detector & N. Or. &
  $\lambda$$_{i}$-$\lambda$$_{f}$ & \AA/pixel \\
 \noalign{\smallskip}
 \tableline
 \noalign{\smallskip}
 1 & 1994/12 & T1KA (1024x1024) & 34 & 3875-9400 & 0.123-0.296 \\
 2 & 1994/05 & T1KA (1024x1024) & 33 & 3875-9000 & 0.124-0.284 \\
 3 & 1993/12 & T1KA (1024x1024) & 34 & 3875-9400 & 0.121-0.288 \\
 4 & 1992/11 & T2KB (2048x2048) & 36 & 3700-9050 & 0.113-0.276 \\
 5 & 1991/09 & TI3 (800x800) & 34 & 3810-8950 & 0.077-0.180 \\
 6 & 1991/05 & TEK2 (512x512) & 32 & 3950-8975 & 0.152-0.310 \\ 
 7 & 1990/10 & RCA1 (512x512) & 40 & 3690-10700& 0.130-0.378 \\
 8 & 1989/12 & TI2 (800x800) & 15 & 7250-9000 & 0.130-0.158 \\ 
 9 & 1989/04 & RCA3 (512x512) & 34 & 3890-9350 & 0.151-0.359 \\
 10 & 1988/09 & RCA3 (512x512) & 33 & 3880-8950 & 0.150-0.344 \\
 11 & 1987/03 & RCA1 (512x512) & 33 & 3880-8950 & 0.151-0.346 \\ 
 \noalign{\smallskip}
 \tableline
 \noalign{\smallskip}
 \end{tabular}
 \end{flushleft}
 \end{table*}

 \normalsize

 \begin{table*}
 \caption[ ]{Lines included in FOE spectral orders in each observing run.
 \label{tab:orders} }
 \begin{flushleft}
 \scriptsize
 \begin{tabular}{l l l l l l l l l l l l l}
 \tableline
 \tableline
 \noalign{\smallskip}
  Or. No. & 1, 2, 3, 10, 11 & 4, 7 & 5 & 6 & 9 & 8 \\ 
 \noalign{\smallskip}
 \tableline
 \noalign{\smallskip}
  1 & Ca~{\sc ii} K & & & Ca~{\sc ii} H & \\
  2 & Ca~{\sc ii} H & & & & \\ 
  3 & & & Ca~{\sc ii} H & H$\delta$ & Ca~{\sc ii} IRT & \\
  4 & H$\delta$ & Ca~{\sc ii} K & & & \\
  5 & & Ca~{\sc ii} H & H$\delta$ & & \\
  6 & & & & H$\gamma$ & \\
  7 & H$\gamma$ & H$\delta$ & & & \\
  8 & & & H$\gamma$ & & \\
  9 & & & & & \\
 10 & & H$\gamma$ & & & Li~{\sc i}\\
 11 & & & & H$\beta$ & H$\alpha$\\
 12 & H$\beta$ & & & & & Ca~{\sc ii} IRT \\
 13 & & & H$\beta$ & & & Ca~{\sc ii} IRT \\
 14 & & & & Mg~{\sc i} b & \\
 15 & Mg~{\sc i} b & H$\beta$ & & & Na~{\sc i} D \\
 16 & & & Mg~{\sc i} b & & \\
 17 & & & & & \\
 18 & & Mg~{\sc i} b & & & \\
 19 & & & & Na~{\sc i} D & \\
 20 & Na~{\sc i} D & & & & Mg~{\sc i} b \\
 21 & & & Na~{\sc i} D & & \\
 22 & & & & & \\
 23 & & Na~{\sc i} D & & H$\alpha$ & H$\beta$ \\
 24 & H$\alpha$ & & & Li~{\sc i} & \\
 25 & Li~{\sc i} & & H$\alpha$ & & \\
 26 & & & Li~{\sc i} & & \\
 27 & & H$\alpha$ & & & \\
 28 & & Li~{\sc i} & & & H$\gamma$ \\
 29 & & & & & \\
 30 & & & & & \\
 31 & & & & Ca~{\sc ii} IRT & H$\delta$ \\
 32 & Ca~{\sc ii} IRT & & & & \\
 33 & & & Ca~{\sc ii} IRT & & Ca~{\sc ii} H \\
 34 & & & & & Ca~{\sc ii} K\\
 35 & & Ca~{\sc ii} IRT & & & \\
 36 & & & & & \\ 
 \noalign{\smallskip}
 \tableline
 \end{tabular}
 \end{flushleft}
 \end{table*}

 \begin{table*}
 \caption[ ]{Stars
 \label{tab:par} }
 \begin{flushleft}
 \scriptsize
 \begin{tabular}{l l l l l l l l l l l l l l}
 \tableline
 \tableline
 \noalign{\smallskip}
  HD & HR & GJ & Name & T$_{\rm sp}$ & MK &
  [Fe/H] & P$_{\rm rot}$ & {\it v}~sin{\it i} & S & Obs. & Pap. \\
  & & & & & &
  (dex) & (days) & (km s$^{-1}$) & & & (I, II) \\
 \noalign{\smallskip}
 \tableline
  {\bf F stars} \\
 \tableline
 \noalign{\smallskip}
  58946 & 2852 & 274~A & $\rho$ Gem & F0~V (SB?) & MK&
  - & - & 68 & - & 8 & \\
  15257 & 717 & - & 12 Tri & F0~III & &
  - & - & 78 & - & 8 & \\
  1457 & - & - & SAO 11104 & F0~Iab & &
  - & - & - & - & 3, 8 & \\
  128167 & 5447 & 557 & $\sigma$ Boo & F2~V & MK &
  -0.387 & - & 7.8 & 0.190 & 11 & \\

  210027 & 8430 & 848 & $\iota$ Peg & F5~V (SB1) & MK &
  -0.079 & - & - & - & 5 & \\
  87141 & 3954 & - & BD+54 1348 & F5~V & &
  0.047 & - & 10 & - & 8 & \\
  55052 & 2706 & - & 48 Gem & F5~III-IV & &
  - & - & 74 & - & 11 & \\
  20902 & 1017 & - & $\alpha$ Per & F5~Ib: & MK* &
  - & - & 18 & - & 8 & \\
  76572 & 3563 & - & 61 Cnc & F6~V & &
  - & - & $<$10 & 0.148 & 11 & \\
  11443 & 544 & 78.1 & $\alpha$ Tri & F6~IV (SB) & &
  0.000 & - & 93 & 0.275 & 5 & \\
  8992 & - & - & SAO 22328 & F6~Ib & &
  - & - & - & - & 3 & \\ 
  187013 & 7534 & 767.1~A & 17 Cyg & F7~V & &
  -0.109& - & 10.0 & 0.154 & 2, 11(2) & I \\ 
  222368 & 8969 & 904 & $\iota$ Psc & F7~V (SB?) & MK &
  -0.127 & - & 5.6 & 0.153 & 4 & \\
  187691 & 7560 & 768.1~A & o Aql & F8~V & & 
  0.059& - & 3.1 & 0.148 & 1, 2, 3, 6(2), 7, 11(2) & I \\
  142373 & 5914 & 602 & $\chi$ Her & F8~V & &
  -0.431& - & 2.4 & 0.147 & 6, 11 & I, II \\
  9826 & 458 & 61 & $\upsilon$ And & F8~V & &
  -0.14 & - & 8 & 0.154 & 5 & II \\
  45067 & 2313 & - & BD-00 1287 & F8~V & &
  -0.16 & - & $<$ 15& 0.141 & 11 & I \\ 
  107213 & 4688 & - & 9 Com & F8~V & &
  0.154& - & 10.0 & 0.135 & 11 & I \\
  122563 & 5270 & - & BD+10 2617 & F8~IV & & 
  -2.74 & - & - & - & 6 & \\
  102870 & 4540 & 449 & $\beta$ Vir & F9~V (SB?) & MK &
  0.180 & - & 4.5 & - & 11 & \\
  22484 & 1101 & 147 & 10 Tau & F9~IV-V (SB?) & &
  -0.106 & - & 2.8 & 0.147 & 4, 8 & \\
 114710 & 4983 & 502 & $\beta$ Com & F9.5~V & MK&
 0.135& 12.35 & 4.3 & 0.201 & 2, 11(2) & I, II \\
 \noalign{\smallskip}
 \tableline
 \end{tabular}
 \end{flushleft}
 \end{table*}
 
 \begin{table*}
 \addtocounter{table}{-1}
 \caption[ ]{Continue}
 \begin{flushleft}
 \scriptsize
 \begin{tabular}{l l l l l l l l l l l l l}
 \tableline
 \tableline
 \noalign{\smallskip}
  HD & HR & GJ & Name & T$_{\rm sp}$ & MK &
  [Fe/H] & P$_{\rm rot}$ & {\it v}~sin{\it i} & S & Obs. & Pap. \\
  & & & & & &
  (dex) & (days) & (km s$^{-1}$) & & & (I, II) \\
 \noalign{\smallskip}
 \tableline
 {\bf G stars} \\
 \tableline
 \noalign{\smallskip}
  39587 & 2047 & 222~AB & $\chi^{1}$ Ori& G0-~V (SB1) & MK &
 -0.084& 5.36 & 8.6 & 0.325 & 10(6) & I, II \\
  143761 & 5968 & 606.2 & $\rho$ CrB & G0+~Va & MK &
 -0.185& - & 5.0 & 0.150 & 11 & I \\
  13974 & 660 & 92 & $\delta$ Tri& G0.5~V (SB2)& MK &
 -0.444& -& 10.0 & 0.232 & 1(2), 3, 4, 5, 10 & I \\
  26630 & 1303 & - & $\mu$ Per & G0~Ib (SB)& MK &
  -0.32 & - & 14 & 0.362 & 8 & \\
  126053 & 5384 & 547 & BD+01 2920 & G1~V & &
  - & - & 1 & 0.165 & 6, 11 & \\
  95128 & 4277 & 407 & 47 UMa & G1-~V & MK &
  0.026 & - & $<$3 & 0.165 & 8 & \\
  67228 & 3176 & - & $\mu^{2}$ Cnc & G1~IVb & &
  0.052 & - & 3.0 & 0.138 & 1(2), 6, 11 & \\ 
  84441 & 3873 & - & $\epsilon$ Leo & G1~II & &
  0.17 & - & $<$17 & - & 9(6), 11(2) & \\

  185758 & 7479 & - & $\alpha$ Sge & G1~II & MK &
  -0.15 & - & 6.0 & - & 2(2) & \\
  - & - & - & Sun & G2~V & &
  0.00 & 25.72 & $<$ 1.7& 0.179 & 1 & I \\
  1835 & 88 & 17.3 & 9 Cet & G2.5~V & MK &
  0.050 & 7.7 & 6 & 0.349 & 1, 3, 4(2), 5 & \\
  221170 & - & - & BD+29 4940 & G2~IV & &
  -1.96 & - & - & 0.106 & 2, 3, 5 & \\
  196755 & 7896 & - & $\kappa$ Del & G2~IV & MK &
  -0.02 & - & 2.7 & 0.152 & 1, 2(2), 3, 4, 5 & \\
  218658 & 8819 & - & $\pi$ Cep & G2~III (SB) & &
  0.01 & - & 22 & 0.237 & 1, 2 & \\
  161239 & 6608 & - & 84 Her & G2~IIIb & MK &
  - & - & 6.0 & 0.138 & 11(2) & \\ 
  11544 & - & - & SAO 22740 & G2~Ib & &
  - & - & - & - & 1, 3 & & \\
  223047 & 9003 & - & $\psi$ And & G3~Ib-II & &
  0.10 & - & $<$ 19 & 0.385 & 8 & \\
  117176 & 5072 & 512.1 & 70 Vir & G4~V & MK &
  -0.035 & - & 1.2 & 0.142 & 6, 11(2) & \\ 
  123 & 5 & 4.1~A & V640 Cas & G5~V & &
  - & - & - & - & 3, 5, 11 & \\
  20630 & 996 & 137 &$\kappa^{1}$ Cet& G5~V (SB?) & MK*&
  0.133& 9.24 & 3.9 & 0.366 & 1(3), 4, 7 & I, II \\
  59058 & - & - & BD+38 1771 & G5~V & &
  - & - & - & - & 8 & \\
  86873 & - & - & BD+32 1970 & G5 & &
  - & - & - & - & 8 & \\
  161797 & 6623 & 695~A & $\mu$ Her A & G5~IV & MK* &
  0.242 & - & 1.2 & 0.136 & 5, 6 & \\
  71369 & 3323 & - & o UMa & G5~III & &
  -0.21 & - & 3.4 & 0.120 & 1 & \\
  - & - & - & $\kappa$ Her & G5~III & &
  - & - & - & - & 2 & \\
  20825 & - & - & 62 Ari & G5~III & &
  -0.14 & - & - & - & 4, 5 & \\
  190360 & 7670 & 777~A & BD+29 3872 & G6~IV+M6~V& &
  0.308 & - & - & 0.146 & 5, 6 & I \\
  221115 & 8923 & - & 70 Peg & G7+~III & &
  -0.03 & - & $<$ 19 & 0.147 & 2 & \\
  101501 & 4496 & 434 & 61 UMa & G8~V & MK* &
  -0.070& 16.68 & 2.3 & 0.311 & 6, 11(2) & I \\
  103095 & 4550 & 451~A & BD+38 2285 & G8~Vp & &
  -1.266 & - & 2.2 & 0.188 & 11 & \\
  188512 & 7602 & 771~A & $\beta$ Aql & G8~IV & MK* &
  -0.30 & - & 1.4 & 0.136 & 1, 2, 4, 5, 7 & I \\
  73593 & 3422 & - & 34 Lyn & G8~IV & &
  - & - & - & 0.117 & 3, 11(3) & \\
  218935 & 8827 & - & 60 Peg & G8~III-IV & &
  - & - & - & 0.120 & 5 & \\
  113226 & 4932 & - & $\epsilon$ Vir & G8~IIIab & MK* &
  0.00 & - & 3.2 & - & 1(3), 2(6), 3, 9(6), 11(6) & \\
  16161 & - & - & $\nu$ Cet & G8~III & &
  -0.38 & - & $<$ 17 & 0.111 & 4, 5 & \\
  104979 & 4608 & - & o Vir & G8~IIIa & MK &
  -0.33 & - & 2.5 & - & 6, 11(2) & \\
  191026 & 7689 & - & 27 Cyg & G8.5~IVa & &
  -0.10 & - & - & - & 4 & \\
  108225 & 4728 & - & 6 Cvn & G9~III & MK &
  -0.11 & - & $<$ 19 & - & 6 & \\
  76294 & 3547 & - & $\zeta$ Hya & G9~IIIa & MK &
  -0.21 & - & - & - & 1, 3 & \\ 
  4128 & 188 & 31 & $\beta$ Cet & G9.5~III & &
  0.13 & - & 4.0 & 0.187 & 10 & \\
 \noalign{\smallskip}
 \tableline
 \end{tabular}
 \end{flushleft}
 \end{table*}
 
 \begin{table*}
 \addtocounter{table}{-1}
 \caption[ ]{Continue}
 \begin{flushleft}
 \scriptsize
 \begin{tabular}{l l l l l l l l l l l l l}
 \tableline
 \tableline
 \noalign{\smallskip}
  HD & HR & GJ & Name & T$_{\rm sp}$ & MK &

  [Fe/H] & P$_{\rm rot}$ & {\it v}~sin{\it i} & S & Obs. & Pap. \\
  & & & & & &
  (dex) & (days) & (km s$^{-1}$) & & & (I, II) \\
 \noalign{\smallskip}
 \tableline
 {\bf K stars} \\
 \tableline
 \noalign{\smallskip}
  185144 & 7462 & 764 & $\sigma$ Dra & K0~V & MK*&
  -0.045& - & 0.6 & 0.215 & 2 & I, II \\
  3651 & 166 & 27 & 54 Psc & K0+~V & MK &
  -9.000& 48.00 & 2.2 & 0.176 & 1, 3, 4, 5, 8 & I \\ 
  198149 & 7957 & 807 & $\eta$ Cep & K0~IV & MK &
  -0.32 & & 0.6 & - & 1, 2 & \\
  6734 & - & - & 29 Cet & K0~IV & &
  -0.25 & - & - & 0.131 & 3 & \\
  168723 & 6869 & 711 & $\eta$ Ser & K0~III-IV & MK &
  -0.42 & - & 2.6 & 0.122 & 6, 11(3) & \\ 
  45410 & 2331 & - & 6 Lyn & K0~III-IV & &
  - & - & - & 0.127 & 11 & I \\
  28 & 3 & - & 33 Psc & K0~III-IV (SB1) & MK &
  -0.31 & - & $<$ 17 & - & 5 & \\
  188947 & 7615 & - & $\eta$ Cyg & K0~III & MK &
  -0.09 & - & 1.8 & 0.103 & 2(2), 8 & \\
  197989 & 7949 & 806.1~A & $\epsilon$ Cyg & K0~III & MK* &
  -0.18 & - & 2.0 & 0.104 & 4(2) & \\
  19476 & 941 & - & $\kappa$ Per & K0~III & MK &
  0.04 & - & $<$ 17 & 0.110 & 5 & \\
  182272 & 7359 & - & BD+33 3434 & K0~III & &
  - & - & - & - & 6 & \\
  19787 & 951 & - & $\delta$ Ari & K0~III & MK &
  -0.03 & - & $<$ 17 & 0.110 & 5 & \\
  8512 & 402 & - & $\theta$ Cet & K0~IIIb & MK &
  -0.22 & - & $<$ 17 & 0.105 & 4, 5 & \\
  12014 & - & - & SAO 22820 & K0~Ib & &
  - & - & - & - & 3 & \\
  10476 & 493 & 68 & 107 Psc & K1~V & MK&
  -0.123& 35.2 & 0.6 & 0.198 & 4(2), 5, 7(2) & I, II \\
  155885 & 6401 & 663~B & 36 Oph B & K1~V & &
  -0.305 & 22.9 & - & 0.384 & 11 & \\
  142091 & 5901 & - & $\kappa$ CrB& K1~IVa & MK &
  -0.04 & - & 0.6 & - & 6, 11 & I \\
  138716 & 5777 & - & 37 Lib & K1~III-IV & MK &
  -0.12 & - & $<$ 19 & - & 11 & \\
  203504 & 8173 & - & 1 Peg & K1~III & &
  -0.14 & - & $<$ 17 & 0.103 & 2(2) & \\
  124897 & 5340 & 541 & $\alpha$ Boo & K1.5~III & MK &
  -0.47 & - & 3.3 & 0.144 & 6(5), 9(3), 11(6) & \\
  6805 & 334 & - & $\eta$ Cet & K2-~III & MK &
  0.04 & - & $<$ 17 & 0.112 & 4 & \\
  210745 & 8465 & - & $\zeta$ Cep & K1.5~Ib & MK &
  0.75 & - & $<$ 17 & 0.293 & 8 & \\
  166620 & 6806 & 706 & BD+38 3095 & K2~V & &
  -0.114 & 42.4 & 0.6 & 0.190 & 1(3), 2, 3, 4(2), 6, 11(2) & I \\
  4628 & 222 & 33 & BD+04 123 & K2~V & &
  -0.235 & 38.5 & - & 0.230 & 4, 5, 10 & I \\
  22049 & 1084 & 144 & $\epsilon$ Eri& K2~V & MK*&
  -0.165 & 11.68 & 2.0 & 0.496 & 5, 7 & I \\
  149661 & 6171 & 631 & 12 Oph & K2~V & &
  -0.004 & 21.3 & 0.6 & 0.339 & 6, 11(2) & \\
  201196 & 8088 & - & BD+15 4340 & K2~IV & &
  - & - & - & - & 1(2), 2, 3(3), 4(2) & \\
  153210 & 6299 & - & $\kappa$ Oph & K2~III & MK*&
  -0.03 & - & $<$ 17 & 0.102 & 6 & \\
  161096 & 6603 & - & $\beta$ Oph & K2~III & MK &
  0.00 & - & 2.5 & 0.103 & 9, 10(4) & \\
  194317 & 7806 & - & 39 Cyg & K2.5~III & MK &
  -0.17 & - & $<$ 19 & 0.148 & 2(2), 4, 8 & \\
  16160 A & 753 & 105~A & BD+06 398 & K3-~V & MK &
  -0.297 & 48.0 & - & 0.226 & 1, 3, 4(3), 5, 7 & I, II \\
  160346 & - & 688 & BD+03 3465 & K3-~V & &
  - & 33.5 & - & 0.300 & 11(3) & \\

  219134 & 8832 & 892 & BD+56 2966 & K3~V & MK&
  -0.017 & - & 2.1 & 0.230 & 8 & I, II \\
  3627 & 165 & - & $\delta$ And & K3~III (SB)& MK &
  0.04 & - & $\leq$ 3 & - & 5 & \\
  136514 & 5710 & - & 6 Ser & K3~III & &
  -0.14 & - & $<$ 17 & - & 6 & \\
  186791 & 7525 & - & $\gamma$ Aql & K3~II & MK &
  -0.29 & - & $<$ 17 & - & 7(3) & \\
  131156 B&5544 B& 566~B & $\xi$ Boo B & K4~V & &
  0.19 & 12.28 & 20 & 1.381 & 11 & I, II \\
  201091 & 8085 & 820~A & 61 Cyg A & K5~V & MK*&
  -0.06 & 35.37 & 0.6 & 0.658 & 1(2), 2(2), 3, 4(6), 5, 8 & I, II \\
  156026 & - & 664 & 36 Oph C & K5~V & &
  -0.279 & 18.0 & 2.2 & 0.770 & 11(2) & \\
  29139 & 1457 & 171.1~A & $\alpha$ Tau & K5+~III & MK &
  -0.16 & - & $<$ 17 & - & 1(5), 3, 4(2), 5(4), 8 & \\
  11800 & - & - & BD+59 363 & K5~Ib & &
  - & - & - & - & 3 & \\
  216946 & 8726 & - & BD+48 3887 & K5~Ib & MK &
  -0.03 & - & - & - & 8 & \\
  201092 & 8086 & 820~B & 61 Cyg B & K7~V & MK &
  -0.10 & 37.84 & 1.4 & 0.986 & 1(2), 2(2), 3, 4(2), 5, 8 & I, II \\
  80493 & 3705 & - & $\alpha$ Lyn & K7~IIIab & MK &
  -0.26 & - & - & - & 8 & \\
 \noalign{\smallskip}
 \tableline
 \end{tabular}
 \end{flushleft}
 \end{table*}
 
 \begin{table*}
 \addtocounter{table}{-1}
 \caption[ ]{Continue}
 \begin{flushleft}
 \scriptsize
 \begin{tabular}{l l l l l l l l l l l l l}
 \tableline
 \tableline
 \noalign{\smallskip}
  HD & HR & GJ & Name & T$_{\rm sp}$ & MK &
  [Fe/H] & P$_{\rm rot}$ & {\it v}~sin{\it i} & S & Obs. & Pap. \\
  & & & & & &
  (dex) & (days) & (km s$^{-1}$) & & & I, II \\
 \noalign{\smallskip}
 \tableline
 {\bf M stars} \\
 \tableline
 \noalign{\smallskip}
  - & - & 906 & V347 & M0~V (K5) & &
  - & - & - & - & 8 & \\
  89758 & 4069 & - & $\mu$ UMa & M0~III (SB) & MK &
  - & - & - & - & 1, 11(2) & \\
  6860 & 337 & 53.3 & $\beta$ And & M0+~IIIa & MK*&
  -0.10 & - & - & 0.319 & 4(2), 5, 8 & \\
  - & - & 4~B & BD+45 4408~B & M0.5~V (K7) & &
  - & - & - & - & 8 & \\
  232979 & - & 172 & BD+52 857 & M0.5~V (K8) & MK&
  - & - & - & 1.909 & 1 & II \\
  1326 A & - & 15~A & GX And & M1.5~V (1) (M2~V)& MK&
  - & - & $<$ 2.9 & - & 1, 2(2), 3, 8 & II \\
  218329 & 8795 & - & 55 Peg & M1~IIIab & MK &
  - & - & - & 0.234 & 1 & \\
  206330 & 8284 & - & 75 Cyg & M1~IIIab & MK &
  - & - & - & - & 8 & \\
  39801 & 2061 & - & $\alpha$ Ori & M1-M2~Ia-Iab & MK*&
  - & - & - & - & 11(2) & \\
  95735 & - & 411 & BD+36 2147 & M2+~Ve (1) & MK &
  -0.20 & - & $<$ 2.9 & 0.424 & 11 & \\
  206936 & 8316 & - & $\mu$ Cep & M2-~Ia & MK*&
  - & - & - & - & 8 & \\ 
  133216 & 5603 & 574.1 & $\sigma$ Lib & M2.5~III & MK &
  - & - & - & - & 11(2) & \\ 
  42995 & 2216 & - & $\eta$ Gem & M2.5~III & &
  - & - & - & - & 11(2) & \\
  2411 & 103 & - & TV Psc & M3~III & &
  - & - & - & 0.211 & 2(2) & \\
  44478 & 2286 & - & $\mu$ Gem & M3~IIIab & MK &
  0.11 & - & - & - & 8 & \\

  14270 & - & - & AD Per & M3~Iab & &
  - & - & - & - & 3 & \\
  - & - & 273 & BD+05 1668 & M3.5~V (1) & &
  - & - & $<$ 2.4 & - & 1(2) & II \\
  55383 & 2717 & - & 51 Gem & M4~IIIab & &
  - & - & - & - & 1, 11(2) & \\
  214665 & 8621 & - & BD+56 2821 & M4+~III & MK &
  - & - & - & 0.259 & 8 & \\
  120323 & 5192 & - & 2 Cen & M4.5~III & MK &
  - & - & - & - & 11 & \\
  130144 & 5512 & - & BD+15 2758 & M5~III & &
  - & - & - & - & 11 & \\
  94705 & 4267 & - & VY Leo & M5.5~III & MK &
  - & - & - & - & 11 & \\
  33664 & 1693 & - & RX Lep & M6~III & &
  - & - & - & - & 11(2) & \\ 
  84748 & 3882 & - & R Leo & M8~IIIe & &
  - & - & - & - & 1(2) & I \\
 \noalign{\smallskip}
 \tableline
 \end{tabular}
 \end{flushleft}

 (1): Henry et al. (1994)
 
 
 SB: Spectroscopic Binary (Duquennoy \& Mayor 1991)

 \end{table*}
 

 \end{document}